\documentclass[12pt,preprint]{aastex}
\shorttitle{Outflow Driven Turbulence}
\shortauthors{Carroll, Frank, Blackman, Cunningham, Tanny}
\providecommand{\e}[1]{\ensuremath{\times 10^{#1}}}
\begin{document}
\title{Isotropically Driven versus  Outflow Driven Turbulence: Observational Consequences for Molecular Clouds}

\author{Jonathan J. Carroll\altaffilmark{1}, Adam Frank\altaffilmark{1}, Eric G. Blackman\altaffilmark{1}}
\altaffiltext{1}{Department of Physics and Astronomy, University of Rochester, Rochester, NY 14620}

\begin{abstract}
Feedback from protostellar outflows can influence the nature of turbulence in star forming regions even if they are not the primary source of velocity dispersion for all scales of molecular clouds. For  the rate and power expected in star forming regions, we previously \citep{Carroll2009} demonstrated that outflows could drive supersonic turbulence at levels consistent with the scaling relations from \cite{matzner2007} although with a steeper velocity power spectrum than expected for an isotropically driven supersonic turbulent cascade. Here we perform higher resolution simulations and combine simulations of outflow driven turbulence with those  of isotropically forced turbulence. We find that the presence of outflows within an ambient isotropically driven turbulent environment produces a knee in the  velocity power spectrum at the outflow scale and a steeper slope at sub-outflow  scales than for a purely isotropically forced case. We also find that the presence of outflows flattens the density spectrum at large scales effectively reducing the formation of large scale turbulent density structures. These effects are qualitatively independent of resolution.  We have also carried out Principal Component Analysis (PCA) for synthetic data from our simulations. We find that PCA as a tool for identifying the driving scale of turbulence has a misleading bias toward low amplitude large scale velocity structures even when they are not necessarily the dominant energy containing scales. This bias is absent for isotropically forced turbulence but manifests strongly for collimated outflow driven turbulence.

 \end{abstract}


\section{Introduction}
\paragraph{}
Supersonic velocity dispersions are a ubiquitous phenomena occuring over a wide range of scales in star forming molecular clouds.  The relationship between velocity dispersions (line-widths) and size was first summarized by \citep{larson}.  While Larson's laws originally applied only to global relationships between different GMC's, \cite{heyer} showed that the same line-width size relationships held within individual GMC's.  These observed velocity dispersions are commonly interpreted as resulting from turbulent gas motion.  The formal distinction between turbulent motion and other sources of velocity dispersion in molecular clouds such as gravitational fragmentation is an important one \citep{field,fbk10},  but for simplicity we presently use the terms turbulence and velocity dispersions interchangeably. 
\paragraph{}
The source of turbulent motions in molecular clouds is a topic of debate.  If self-gravity is ignored then large scale sources of energy for turbulence include the kinetic energy injection from supernova and galactic differential rotation. Sources of turbulent energy on smaller scales include stellar feedback in the form of HII regions, radiatively driven winds or accretion driven outflows.   This raises the issue of the so-called injection scale, and whether there are in fact more than one dynamically important injection scale where energy is deposited into the turbulent spectrum.  For a clump scale cloud ($\le  10$pc),  theoretical scenarios tend to divide the sources of turbulent forcing mentioned above into external and internal:  External forcing models consider the injection scale to be larger than the cloud scale (e.g. Supernova driven ISM turbulence) whereas internal forcing would refer to injection scales smaller than the cloud size (HII regions, outflows)\citep{McKeeOstriker} or winds from B stars \citep{Valverde2010}.  Note that if self-gravity is included, then gravitational contraction from very large scales can itself produce velocity dispersions that are consistent with Larson's laws \citep{field,fbk10}. That represents an alternative paradigm for driving turbulent-like motions in molecular clouds. Whatever the source of the motions, in order for the turbulence in the parsec and subparsec scale clumps and cores to remain in an approximate steady state, it needs to be continually driven over time scales longer than a crossing time either internally via stellar feedback or externally via a cascade down from the ISM.
\paragraph{}
Distinct from the question of the origin of turbulent motions, is the role of turbulent 'feedback' on the star formation process.  In the absence of turbulent support, most of the mass within a given structure (GMC, clump, core) would collapse into stars within one free fall time giving an efficiency per free fall time $\epsilon_{ff}=1$ \citep{McKeeOstriker}.  Observations however yield surprising low values of $\epsilon_{ff}$ ranging from .01 to .1 not only for GMC's but also for substructures within GMC's of mean densities $n_H$ ranging from $10^2$ to $10^5$ cm$^{-3}$ \citep{krumholz-turb-reg,krumholztan}.  One explanation for these low values of $\epsilon_{ff}$ is the that the observed turbulent motions are signatures of internal kinetic support against collapse. We emphasize that the global energy budget of turbulence is a separate issue from the the issue of local turbulent feedback. For example, star clusters form on scales that are a fraction of the total molecular cloud scale $L_{cloud}/L_{cluster} \sim 10$ and dynamical feedback from outflows can impede star formation on cluster forming scales even if the total energy budget from outflows is insufficient to power the observed turbulent energy budget of the entire GMC.  This means that internal sources of turbulent forcing may be important on small scales whether or not they are important on large scales.  On even smaller scales (ie. $L_{core} \sim .1$ pc) outflows are likely responsible for the dispersal of gas \citep{VelusamyLanger1998, MatznerMckee2000, Myers2008} especially if the outflow opening angle widens with time \citep{ArceSargent2006}, although the complex interaction between cores and their outflows requires more numerical modeling.  In addition, not all forcings are equal: some forcings are isotropic and some are not. This distinction has observational consequences.  
\paragraph{}
In this paper we focus on the physics and observational consequences of collimated outflow driving into both initially laminar and initially turbulent ambient environments.  Theoretical treatments of feedback from multiple spherical outflows creating a self-regulating star forming system was first explored by \cite{normansilk}.  Analytical work by Matzner \citep{matzner00,matzner01,matzner2007} has explored the role of collimated outflow feedback on clouds. \cite{krumholz} also considered the nature of feedback via outflows, concluding that these systems provide an important source of internal driving in dense star forming cores. The first simulation based study of the problem was presented by \cite{maclow-outflows}.  More recently \cite{linaka} and \cite{nakamura} have mapped out the complex interplay between star formation outflows and turbulence.  Their simulations include a self-consistent formulation of driving outflows from newly formed stars and they concluded that protostellar outflows were a viable means of generating turbulence in star forming clusters. \cite{Wang2010} built on these studies using higher resolution AMR studies demonstrating the vigorous feedback that outflows are able to provide a cluster including the significant modification of the star formation efficiency and development of massive stars.  We note that the Nakamura, Li and Wang studies all focused on cluster scales rather than molecular cloud scales. In addition they did not focus specifically on the mechanisms by which outflows generated turbulence or the physics of power spectra signatures.  We also note a study of single jets by \cite{banerjee} that explored the volume filling averages of post shock concluding that single jets would not leave enough supersonic material in their wakes to act as a relevant source of internal forcing.
\paragraph{}
The study of the detailed mechanics of outflow driven turbulence was taken up in a series of papers by \cite{cunningham-jc,cunningham-cavity,cunningham-sj}. These investigations built off observational work by \cite{quillen} for NGC 1333 which showed that turbulence was likely driven not by active outflows but by the interaction of fossil shells driven by extinct outflows.  In \cite{cunningham-sj} it was shown how single outflows in a turbulent environment would be disrupted leading to their energy resupplying turbulent motions. In \cite{Carroll2009} a study of multiple, randomly oriented jets driven into a quiescent medium was undertaken. The results of these simulations showed conclusively that interacting outflows can lead to turbulent motions.  Explorations of the energy spectrum and the density PDF showed that statistically steady isotropic random motions were maintained by the outflows.  Of particular interest was the fact that an injection scale associated with outflow cavity interaction $K_{outflow}$ was apparent in the energy spectrum $E(k)$.  Also the power law index of the spectrum was steeper than would be expected for Burgers type turbulence such as that generated when less realistic isotropic forcing at large scales is used. \cite{Carroll2009} found that the steeper spectrum was associated with outflow cavities sweeping up vortices at smaller scales.  {\it Thus the Carroll results provided both proof that outflows can drive turbulence and provided some insights into the physics involved and its statistical signatures}.
\paragraph{}
The goal of discerning scales of  turbulent driving in real molecular clouds has, recently, received considerable attention.  In \cite{bruntheyermaclow} PCA was used to determine the dominant scales of turbulent driving based on molecular line emission.  Based on comparisons of  first and second eigenimages of molecular channel maps for $^{13}$CO and $^{12}$CO they find that the data support the conclusion that that the bulk of the energy driving for turbulence comes from scales much larger than that of clumps.  \cite{padoan-vca} reached a similar conclusion by applying VCS (Velocity Coordinate Spectrum) \citep{lazarian09} on $^{13}$CO data.   \cite{bruntheyermaclow} also analyzed C$^{18}$0 data which probes denser regions of the clumps, and this data did reveal evidence for a second, subparsec  driving scale consistent with that of outflows.  While  neither of these studies preclude the importance of feedback from outflows on very small scales, neither of these papers actually invoke full 3-D simulations of collimated outflow driven turbulence in their conclusions about driving scale.  Thus there is  a potential loophole in the application of methods that require a density-velocity scaling relation in order to interpret observations as a signature of kinetic energy density on a given scale. One of the topics we address in this paper is the bias that results from this specific issue.
\paragraph{}
In section 2 we discuss the basic parameters and setup of our simulations and choices of forcing for both our collimated outflow driving and our isotropically forced cases.  In section 3 we discuss the results comparing the isotropically driven turbulence with collimated outflow driven turbulence and the interplay between the two. In particular, we analyze the growth and saturation of turbulence, the velocity and density spectra, the characteristic driving scales, and the relative insensitivity to resolution. In section 4 we uncover biases in the current use of PCA  to infer the scales of turbulent driving when comparing isotropically forced to outflow driven turbulence.  We show that PCA fails to identify where the dominant driving power is when comparing sources of turbulence with different forcings.  We conclude in section 5.
\section{Numerical Model} \label{numerical_model}
As in \cite{Carroll2009} we used an MHD code called AstroCUB but without magnetic fields for this work.  The code is 2nd order accurate in space and time and uses a non-split CTU method with upwinded CT as described by \cite{GardinerStone2008}.  All four of the simulations were performed on a periodic cube of length $L_{box}=1.48$ pc initialized with a uniform density $\rho_{_0}=2.51\e{-20}$g cm$^{-3}$ and mean particle weight of 2.1 amu at 10 K giving a sound speed of .20 km/s. Each of the simulations was driven by some combination of outflow feedback (\S \ref{Outflow_Feedback}) and/or isotropic forcing (\S \ref{Isotropic_Forcing}) described below.  Run HDO was forced solely via outflows, run HDI was forced isotropically, and run HDOI employed both forms of forcing each of comparable strength.
It is worth mentioning that the difference in velocity between typical outflows (240 km/s) and the turbulence they generate (1 km/s) makes simulations that model both types of flows computationally intensive.  This is because with Eulerian grid-based codes, the computational time steps are limited by the highest speeds (240 km/s) but the timescale of interest is set by the turbulent velocity (1 km/s).  Therefore since simulations with outflow forcing would normally take of order 100 times as long as those with just isotropic forcing, we started runs with outflow forcing at lower resolutions and then periodically doubled the resolution until the desired resolution was reached.  Since supersonic turbulence decays on the order of 1 dynamical time $t_d=\frac{L_{box}}{v_{rms}}$, any effects from regridding are expected to also disappear after $1 t_d$.  For this reason, all of the spectra were taken after at least $1 t_d$ after the final regridding.  All of the runs and the regridding times are summarized in Table \ref{tableofruns}.  
\begin{deluxetable}{cccccccc} 
\tablecolumns{8} 
\tablewidth{0pt} 
\tablecaption{Table of Runs \label {tableofruns}} 
\tablehead{ 
\colhead{Run}  & \colhead{Isotropic}    & \colhead{Outflow}  & \colhead{Initial} & \colhead{Final} & \multicolumn{3}{c}{Regrid Time in $\mathcal{T}$} \\
\colhead{} & \colhead{Forcing} & \colhead{Forcing} & \colhead{Resolution} & \colhead{Resolution} & \colhead{128} & \colhead{256} & \colhead{512} }
\startdata 
HDO & No & Yes & $128^3$ & $512^3$ & \nodata & 5 & 6 \\
HDI & Yes & No & $512^3$ & $512^3$ & \nodata & \nodata & \nodata\\
HDOI & Yes & Yes & $64^3$ & $512^3$ & 3 & 5 & 6\\
\enddata 
\end{deluxetable} 
\subsection{Outflow Feedback} \label{Outflow_Feedback}
During the runs with outflow feedback, highly collimated bipolar outflows of momentum $\mathcal{P}=20.0 M_{\sun}$ km/s were launched at the volumetric rate of $\mathcal{S}=58.4 $pc$^{-3}$Myr$^{-1}$ or with a period of $T_{launch}=\mathcal{S}L_{box}^3=5.28$ kyr.  Combining the outflow momentum $\mathcal{P}$ and rate $\mathcal{S}$ with the mean density $\rho_{_0}$ determines the characteristic outflow scales of mass, length, and time \citep{matzner2007}:
\begin{displaymath}
\begin{array}{lcr}
\mathcal{M}=\frac{\rho_{_0}^{4/7}\mathcal{P}^{3/7}}{\mathcal{S}^{3/7}} &
\mathcal{L}=\frac{\mathcal{P}^{1/7}}{\rho_{_0}^{1/7}\mathcal{S}^{1/7}} &
\mathcal{T}=\frac{\rho_{_0}^{3/7}}{\mathcal{P}^{3/7}\mathcal{S}^{4/7}} \\
\end{array}
\end{displaymath}
Combining these gives other characteristic quantities.  Of particular interest is the characteristic velocity, the characteristic wave number, and the characteristic accleration:
\begin{displaymath}
\begin{array}{lll}
\mathcal{V}=\frac{\mathcal{L}}{\mathcal{T}} &
\mathcal{K}=\frac{2\pi}{\mathcal{L}} &
\mathcal{A}=\frac{\mathcal{L}}{\mathcal{T}^2} \\
\end{array}
\end{displaymath}
Table \ref{Outflow_Scales} summarizes the various resulting outflow scales.  
\begin{deluxetable}{llll}
\tablecolumns{4}
\tablewidth{0pt}
\tablecaption{Outflow Scales \label{Outflow_Scales}}
\tablehead{
\colhead{} &  \colhead{cgs units} & \colhead{astronomical units} & \colhead{isotropic scale}}
\startdata 
$\rho_{_0}$ & $2.51\e{-20}$ g cm$^{-3}$ & $371$ M$_\sun$ pc$^{-3}$ & $\rho_{_0}$ \\
$\mathcal{P}$ & $3.98\e{39}$ g cm s$^{-1}$ & $20.0$ M$_\sun$ km/s &  $128 \mathcal{P}$\\
$\mathcal{S}$ & $6.31\e{-68}$ cm$^{-3}$ s$^{-1}$&  $58.4 $pc$^{-3}$ Myr$^{-1}$ & $\mathcal{S}/128$\\
$\mathcal{M}$ & $3.73\e{34}$ g & 18.7 M$_\sun$ & $64 \mathcal{M}$\\
$\mathcal{L}$ & $1.14\e{18}$ cm & $.370$ pc & $4\mathcal{L}$\\
$\mathcal{T}$ & $1.07\e{13}$ s & $.338$ Myr & $2\mathcal{T}$\\
$\mathcal{V}$ & $1.07\e{5}$ cm/s & $1.07$ km/s & $2\mathcal{V}$\\
$\mathcal{A}$ & $1.00\e{-8}$ cm/s$^2$ & $3.23$ pc/Myr$^2$ & $\mathcal{A}$ 
\enddata 
\end{deluxetable}

\paragraph{}
Each outflow was launched from a cylindrical source region randomly located and oriented with radius $r_o$ and length $2z_o$.  Instead of constantly setting the values of density and momentum within each launch region as in \cite{Carroll2009}, source terms were calculated to supply the mass and momentum needed to maintain the desired density and velocity profiles. The density profile was chosen to be constant $(\rho_{_0})$ while the velocity profile contained a quiescent core $(|z|<z_i)$ with $|z|$ the direction of the outflow.  Beginning with bipolar acceleration regions from $z_i < |z| < z_o$ the profile is described below.
\begin{displaymath}
v_{z} = \left\{
  \begin{array}{lr}
    0 & : 0 \leq |z| < z_{i} \\
    \mbox{sign}\left(z\right)\frac{V_o(|z|-z_{i})}{\Delta z}  & : z_{i} \leq |z| \leq z_o \\
  \end{array}
 \right .
\end{displaymath}
The source terms for density $S_\rho$ and momentum $S_{\rho\mathbf{v}}$ were calculated by substituting the launch profiles into the conservation equations and solving for a steady state solution.
\begin{equation}
\frac{\partial \rho}{\partial t} = - \nabla \cdot (\rho \mathbf{v}) + S_{\rho} = 0 \Rightarrow S_{\rho}=\nabla \cdot (\rho \mathbf{v})
\end{equation}
\begin{equation}
\frac{\partial \rho \mathbf{v}}{\partial t} = - \nabla \cdot (\mathbf{v} : (\rho \mathbf{v})) - \nabla P + S_{\rho\mathbf{v}} = 0 \Rightarrow S_{\rho\mathbf{v}} =  \nabla \cdot (\mathbf{v} \colon (\rho \mathbf{v})) + \nabla P 
\end{equation}
As was mentioned before, modeling outflows and the turbulence they produce is computationally difficult due to the high ratio between the outflow velocity $V_o$ and the turbulent velocities they produce $\mathcal{V}$.  Fortunately, since outflows are highly radiative, it is their momentum which is most critical in modeling their feedback.  This allows us to increase the mass injected while decreasing the velocity thereby reducing the ratio $V_o/\mathcal{V}$.  So while the momentum injected by each outflow was modeled on a $.5 M_\sun$ star ejecting $1/6$ of its mass at $240$ km/s, the actual mass injected by each outflow was increased from $.083 M_\sun$ to $.305 M_\sun$ to allow for a slower velocity jet ($V_o/\mathcal{V}=61$ instead of $224$).  This outflow mass loss increases the mean density by $\approx 2\%$ for each outflow time $\mathcal{T}$.  Since the simulations were run for $8 \mathcal{T}$ the overall increase in the mean density was $<16\%$.  In addition the mass loss rate was chosen to be $1.30\e{-4}M_\sun$/yr giving an outflow duration of $2.35$ kyr.  The actual duration of the outflow was modified slightly to allow for a gradual ramp down of the velocity while keeping the total momentum injected equal to $\mathcal{P}$.  This helped to prevent a large rarefaction from developing once the outflow sourced terms were shut off.  Since the outflow period was $T_{launch}=5.28$ kyr, only one outflow was ever active at a given time.  Table \ref{outflow_params} summarizes the outflow parameters.
\begin{deluxetable}{llll}
\tablecolumns{4}
\tablewidth{0pt}
\tablecaption{Outflow Params \label{outflow_params}}
\tablehead{\colhead{description} & \colhead{symbol} & \colhead{value} & \colhead{comment}}
\startdata
density & $\rho_{_0}$ & $2.5\e{-20}$ g/cm$^3$ & \nodata{}\\
velocity & $V_o$ & $65.5$ km/s & \nodata{} \\
duration & $t_o$ & $2.34$ kyr & \nodata{}\\
period & $T_o$ & $5.28$ kyr & \nodata{}\\
radius & $r_o$ & $5960$ AU & $10\Delta x$ at $512^3$ \\
inner buffer & $z_i$ & $2\Delta x$ & $1190$ AU at $512^3$ \\
launch thickness & $\Delta z$ & $3580$ AU & $6\Delta x$ at $512^3$\\
opening angle & $\theta_o$ & 0 \degr & \nodata{} \\
mass loss rate & $\dot{M}$ & $1.30\e{-4}M_\sun$ yr$^{-1}$ & \nodata{} \\
\enddata
\end{deluxetable} 
\subsection{Isotropic Forcing} \label{Isotropic_Forcing}
In addition to forcing via outflow feedback, a constant isotropic solenoidal forcing at the box scale $L_{box}=4\mathcal{L}$ was used to mimic the cascade of turbulent energy from larger scales.  The mean acceleration of the isotropic forcing was $\left<\mathbf{a}\right>=.95 \mathcal{A}$ while the components and phases of each wavevector were chosen randomly.  Table \ref{isotropic_params} lists the forcing vectors and phases for each wave mode.  The total force applied to each cell was 
\begin{equation}
\vec{f}=\rho\vec{a}\left(\vec{x}\right)=\rho\sum_{n}{\vec{A_n}\cos{\left(\vec{k_n}\cdot\vec{x}+\phi_n\right)}}
\end{equation}
\begin{deluxetable}{lll}
\tablecolumns{3}
\tablewidth{0pt}
\tablecaption{Isotropic forcing components \label{isotropic_params}}
\tablehead{\colhead{$\vec{k}$} & \colhead{$\vec{A}$} & \colhead{$\phi$}}
\startdata
$[1,0,0]$ & $[      0, .04087,  .81547]$ & 1.6519 \\
$[0,1,0]$ & $[-.64450,      0, -.50128]$ & 1.1919 \\
$[0,0,1]$ & $[-.41577, .70271,       0]$ & 3.7784 \\
\enddata
\end{deluxetable}
In a similar manner to the outflow forcing, we can combine the isotropic forcing scale $L_{box}$ with the mean acceleration $\mathcal{A}$ and the ambient density $\rho_{_0}$ to calculate characteristic scales for the isotropically driven turbulence.  Of particular interest is the velocity scale $\mathcal{V}_I=\sqrt{L_{box}\mathcal{A}}=\sqrt{4\mathcal{L}\mathcal{A}}=2\mathcal{V}$ and the associated time scale $\mathcal{T}_I=\mathcal{V}_I/\mathcal{A}=2\mathcal{T}$.  While both types of forcing have the same mean accleration and inject momentum at comparable rates, the isotropic forcing has a greater velocity scale simply because it injects energy on a larger spatial scale.  For completeness, all of the equivalent scales associated with the isotropic forcing are listed in table \ref{Outflow_Scales}
\section{Results} \label{results}
\subsection{Growth and Saturation of Turbulence} \label{growth}
\begin{figure}
\plotone{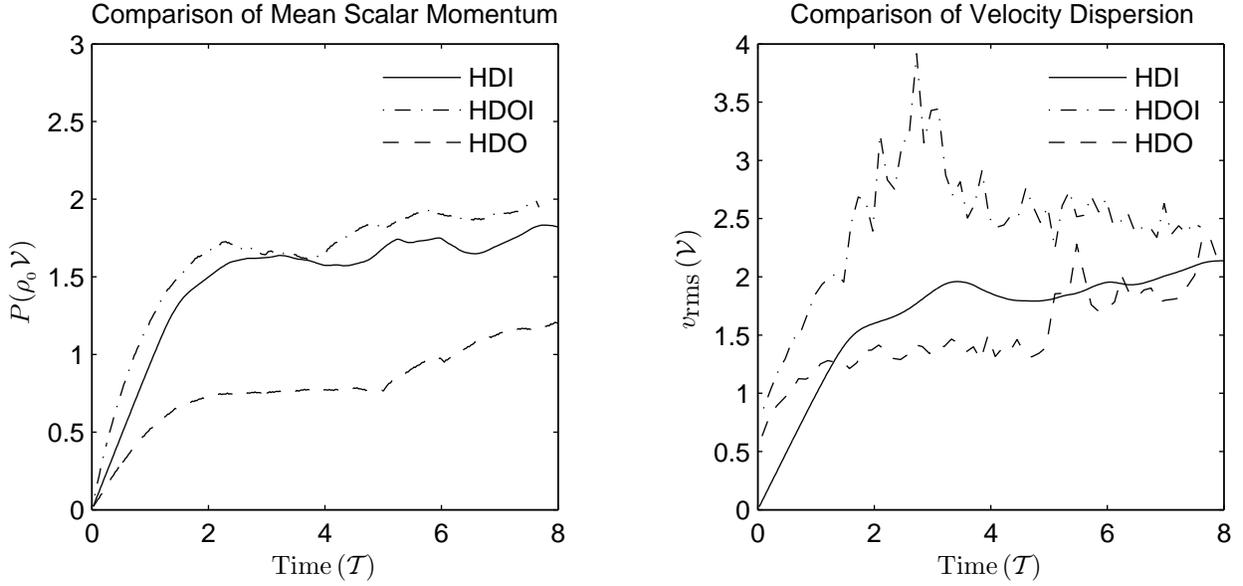}
\caption{Time series showing development of the mean scalar momentum density $P$ and the velocity dispersion $v_{rms}$ for runs HDI (solid lines), HDOI (dash-dotted lines), and HDO (dashed lines).  All quantities are scaled to the outflow scales (ie $\mathcal{T}$,$\mathcal{V}$, \& $\rho_{_0}$).}
\label{totals}
\end{figure}

\begin{figure}
\plotone{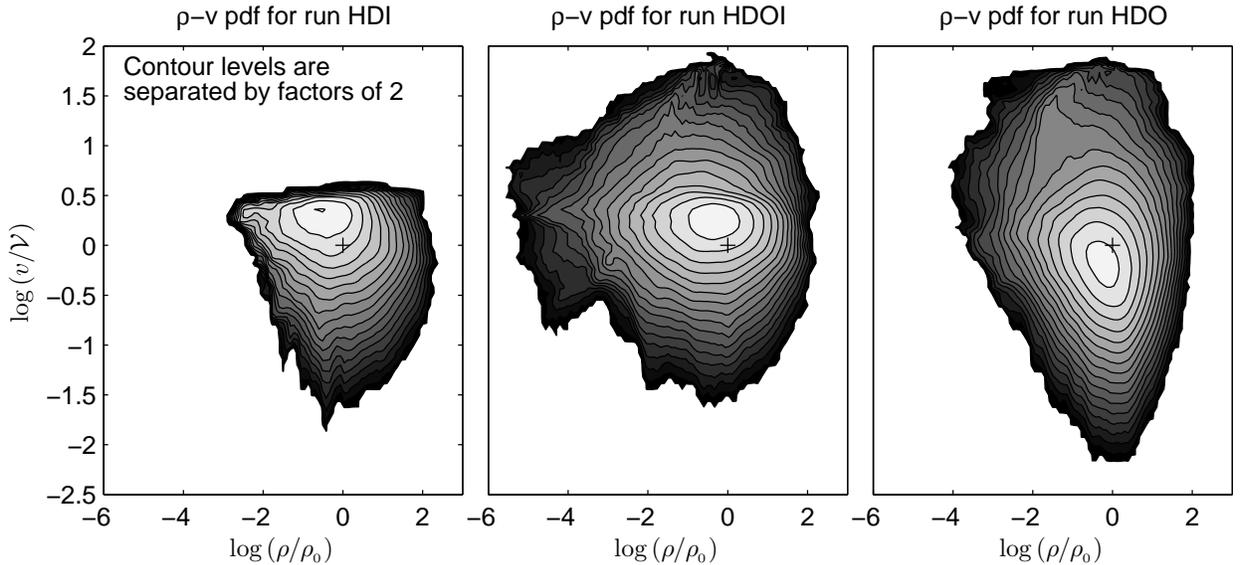}
\caption{Joint probability distribution functions of velocity (y-axis) and density (x-axis) for runs HDI (left), HDOI (middle), and HDO (right).  Lighter areas correspond to higher probability.  The crosshairs are at the mean density $\rho_{_0}$ and the outflow scale $\mathcal{V}$ }
\label{2Dpdf}
\end{figure}

Figure \ref{totals} shows the growth and saturation of the mean scalar momentum $P\equiv <\rho|\vec{v}|>$ and the velocity dispersion $v_{\mbox{rms}} \equiv \sqrt{<v^2>}$ with just isotropic forcing (HDI), just outflow forcing (HDO), and where both types of forcing are present (HDOI).  Since both types of forcing have approximately the same mean acceleration ($\mathcal{A}$ for the outflow forcing, and $.95\mathcal{A}$ for the isotropic forcing), the initial growth of the scalar momentum seen in the left panel of figure \ref{totals} for run HDOI is larger (about twice) that of runs HDO and HDI as expected, although due to the lower initial resolutions of run HDO and HDOI, the initial growth in the outflow momentum injection rate appears to be $\approx 80\% \rho_{_0}\mathcal{V}/\mathcal{T}$.  The effects of the resolution can also be clearly seen in the resaturation of the scalar momentum for run HDO following regridding at $t=5\mathcal{T} \mbox{ and } 6\mathcal{T}$ and somewhat less so for run HDOI at $t=3\mathcal{T},5\mathcal{T},\mbox{ and }6\mathcal{T}$.  Each time the resolution is doubled, the dissipation scale is halved, the dissipation rate is reduced, while the injection rate remains the same - leading to a higher saturation level. 
\paragraph{}
Even though both types of forcing have the same mean acceleration, the final saturation levels for the mean scalar momentum for runs HDI and HDO differ due to the difference in length scales associated with the driving.  The isotropic forcing injects its energy/momentum at a larger scale ($L_{box}=4\mathcal{L} $ instead of $\mathcal{L}$), leading to a longer turnover or cascade time ($2\mathcal{T}$ instead of $\mathcal{T}$), and a slower dissipation.  This results in a higher saturation level consistent with the velocity scale for the isotropic forcing ($\mathcal{V}_I=2\mathcal{V}$).  The saturation level for run HDOI is slightly larger then that of run HDI due to the additional momentum injected by the outflows.  Given the scaling relations, we might expect the velocity dispersion for runs HDI and HDO to follow a similar trend.  Instead, we see that in spite of having a lower turbulent velocity scale, the outflow driven turbulence has a comparable velocity dispersion.  This can be explained by considering the high velocity nature of the outflows themselves ($V_0/\mathcal{V}=61$).  It takes very little high velocity gas to significantly contribute to the velocity dispersion, and when outflows are present there is plenty of it around.  In addition, the high velocity impulsive nature of the outflow driving leads to the high degree of variability in the velocity dispersion.
\paragraph{}
In figure \ref{2Dpdf} we show joint probablility distribution for the three hydodynamic runs.  The isotropically forced runs shows the expected log-normal distribution with density.  As was demonstrated in \cite{Carroll2009} outflows will also produce such a log-normal distribution in $\rho$ as can be seen in the higher resolution HDO runs.  Thus we confirm the conclusions that transient outflow cavities can set the bulk of initially quiescent material into random but statistically steady supersonic motions.  The pdf's for runs HDO and HDOI also reveal the presence of the high velocity material produced by the outflows responsible for the high velocity dispersions and the higher degree of variability.  The isotropic forcing on the other hand is only able to accelerate material at $\mathcal{A}$ over a distance $L_{box}$ to velocities of $\approx \sqrt{2L_{box}\mathcal{A}}=\sqrt{8}\mathcal{V}$ leading to the cutoff in the left panel of figure \ref{2Dpdf} (run HDO) and the horizontal kink in the middle panel (run HDOI).
\subsection{Driving Scales of Turbulence}
\begin{figure}
\plotone{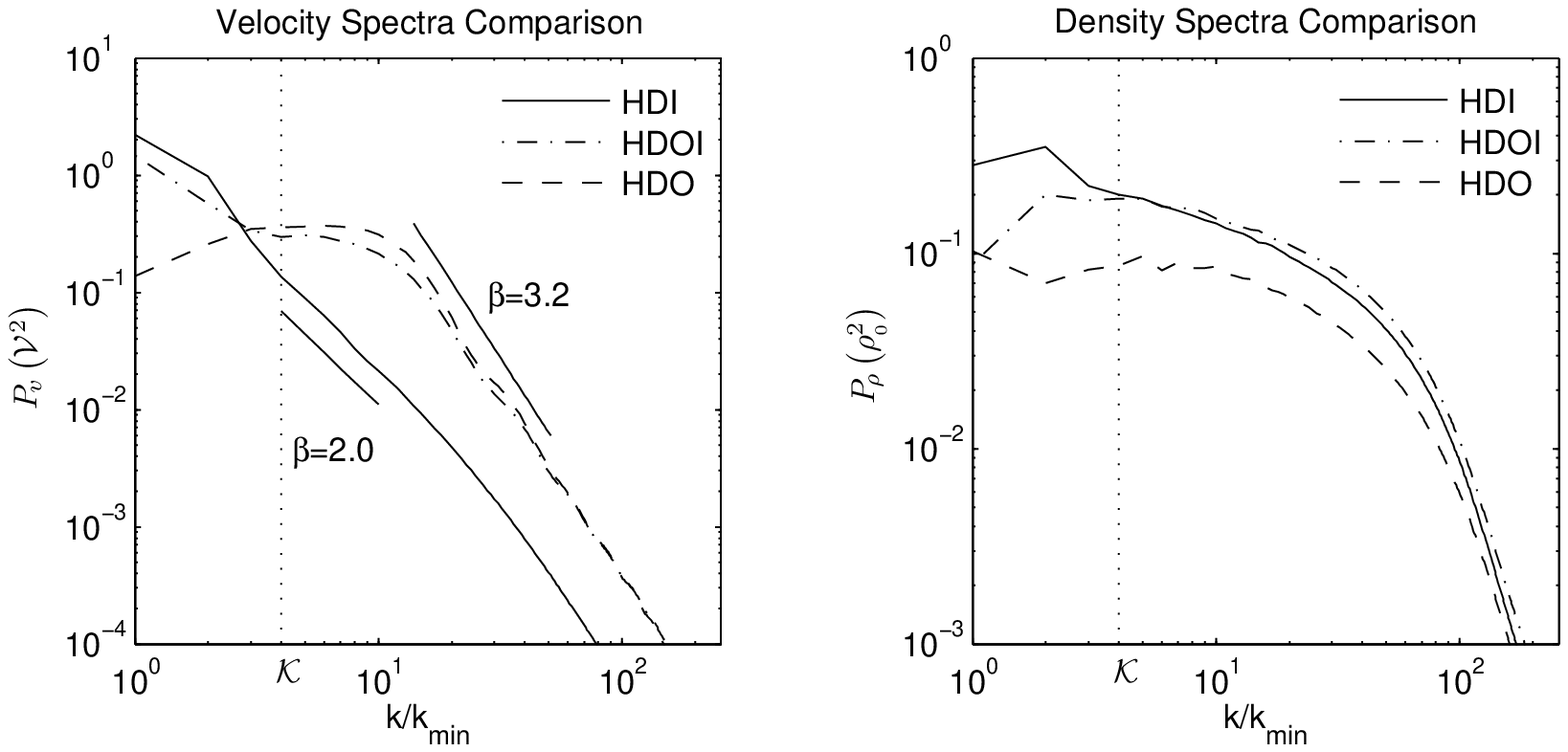}
\caption{Velocity power spectrum (left) and density spectrum (right) for runs HDI (solid line), HDOI (dash-dotted line), and HDO (dashed line).  The vertical dashed line corresponds to the outflow wave number $\mathcal{K}$ and all quantities are scaled to the outflow scales.}
\label{force_comp}
\end{figure}

  In order to better understand how these two types of forcing influence their environments over a range of scales, we looked at the velocity and density spectra for runs HDO, HDI, and HDOI after a statistically steady state had been reached at a resolution of $512^3$.  First, as would be expected, the driving scales associated with the outflow forcing ($\mathcal{K}$) and the isotropic forcing ($\frac{k}{k_{min}}=1$) are clearly evident in the velocity spectra seen in figure \ref{force_comp}.  The velocity spectra for runs HDO and HDOI have a clear peak and knee respectively at the outflow wave number $\mathcal{K}$, while the velocity spectra for runs HDI and HDOI peak at the isotropic forcing scale $\frac{k}{k_{min}}=1$.  Secondly as would also be expected, run HDI has a spectral index of $\beta=2.0$ over the marginally resolved inertial range, consistent with previous simulations of supersonic, isothermal, isotropically-forced turbulence as well as supersonic cascade and shock-dominated models.  Somewhat unexpectedly, however, is the steep slope ($\beta=3.2$) present in the velocity spectra for runs HDO and HDOI at sub-outflow scales.  This steep slope along with the significantly greater amount of energy imply more coherent higher-velocity structures on sub-outflow scales.  This is reasonable however, given the nature of the outflow forcing.  Outflow cavities are able to expand coherently out to distances $\mathcal{L}$ before they have decellerated to the background turbulent velocities, at which point they lose coherence.  As the cavities break up they will stir motions with velocities $\mathcal{V}$ over size scales $\mathcal{L}$ and inject their remaining momentum into turbulent motions.  The time scale for these turbulent motions to cascade however $\frac{\mathcal{L}}{\mathcal{V}}=\mathcal{T}$, is the same as the timescale between successive outflows, so each outflow effectively destroys any decoherent motions left by its predecessors.  It is interesting that the source of the turbulent energy is also, to some degree, the source of its dissipation.  We conjecture that this conclusion will hold for all forms of internal driving where stellar sources feed energy back into the parent cloud.  Our results highlight the difference between simulations which track the more realistic discrete driving and those which impose an apriori driving spectrum at large scales to model an unknown process of external driving.
\paragraph{}
In contrast to the effect outflows have on the velocity spectra (a steepening at sub-outflow scales), the density spectra undergoes a flattenning at supra-outflow scales as seen in the right panel of figure \ref{force_comp}.  Note the density spectrum for run HDO is quite flat at large scales and that run HDOI is flattened compared to run HDI.  This is not too surprising when one considers that each outflow is randomly placed and oriented.  To gain a visual sense of what this means we present in the top panels of figure \ref{log_density} 2D cuts of density through both the HDOI and HDI simulations once saturation has been achieved.  In addition, in the bottom panels of figure \ref{log_density} we show log column density maps to give a sense of the difference in observational appearance the two modes of turbulent driving would obtain.  
\paragraph{}
When only outflows are present (as in run HDO), there is no source for large (supra-outflow) scale coherent motions producing large scale coherent density structures.  The random placement and orientation of outflows produces fairly random density structures and a fairly white spectra.  When only large scale isotropic forcing is present (as in run HDI) do large scale motions driven by large scale isotropic forcing generate large scale coherent density structures.  These can be clearly seen in the right panels of figures \ref{log_density}.  The time scale however needed for these structures to form is $\mathcal{T}_I=2\mathcal{T}$.  Thus when both large scale forcing and outflows are present (as in run HDOI), these large scale density structrues are unable to survive being torn apart by sequential rounds of outflow generation (which occur on a timescale $\mathcal{T}$).  Thus outflows suppress small $k$ modes in density resulting in a flatter density spectra as seen in figure \ref{force_comp} as well as a less coherent density field as seen in the left panels of figure \ref{log_density}.  This disruption in large scale density structures is somewhat mirrored in the disruption of large scale velocity structures.  This explains the decrease in power at large scales seen in run HDOI as compared to run HDI.
\subsection{Resolution study} \label{Resolution}
\begin{figure}
\plotone{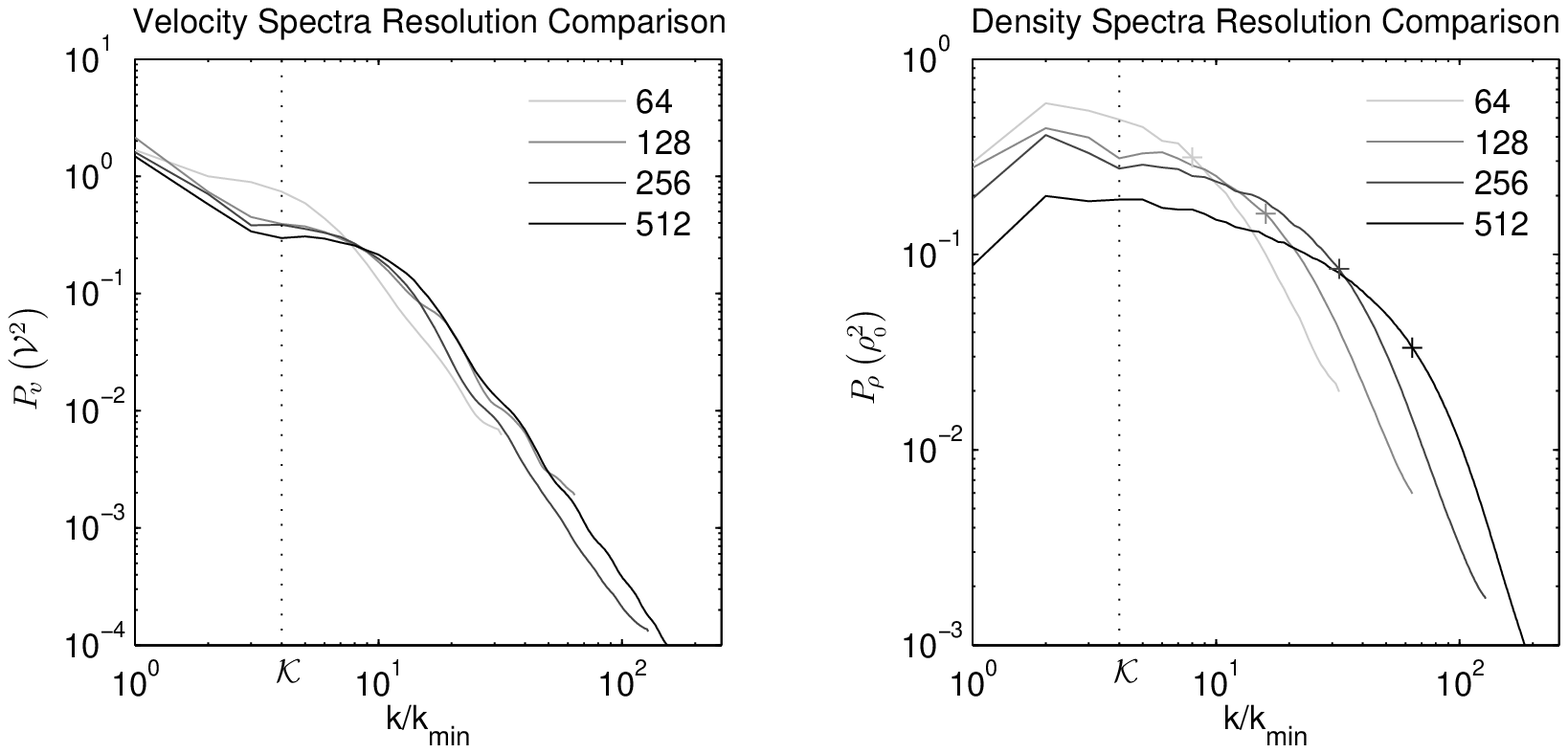}
\caption{Plots showing velocity and density power spectrum for run HDOI at different resolutions.  Note the insensitivity of the velocity spectrum to resolution at sub-outflow scales.  Also note the strong resolution-dependent steepening of the density spectra around $8\Delta x$ marked by the '+'s in the right panel.  The vertical dashed line corresponds to the outflow wave number $\mathcal{K}$ and all quantities are scaled to the outflow scales.}
\label{res_comp}
\end{figure}

To better understand the role numerical dissipation might play in the density and velocity spectra, we completed run HDOI at $64^3, 128^3, 256^3$ as well as $512^3$ in order to perform a resolution study.   Figure \ref{res_comp} shows both the velocity and density power spectra at $\sim 8 \mathcal{T}$ at resolutions of $64^3$, $128^3$, $256^3$, and $512^3$.  The velocity spectra are fairly insensitive to numerical resolution, especially at sub-outflow scales.  In cascade models of turbulence, the lack of turbulent eddies near the dissipation scale reduces the rate at which energy can cascade creating a pileup of energy at scales above the dissipation scale.  This bottleneck and subsequent steepening depends entirely on the dissipation scale and would be resolution dependent.  The lack of any resolution dependent bottleneck effect is however consistent with the lack of any cascade at sub outflow scales, owing to outflows themselves being responsible for the turbulent dissipation.  The outflows themselves create a kind of bottleneck, but this bottleneck is at the outflow scale - not the dissipation scale - and is well resolved (except perhaps marginally so at $64^3$).
\paragraph{}
The density spectra, on the other hand, due show strong resolution dependent effects.  At supra-outflow scales, the spectra become flatter as the resolution increases due to the ability of the outflows to more finely shred the large scale density structures.  The spectra also appear to steepen at a resolution dependent scale of $\sim 8\Delta x$ most likely due to numerical diffusion near shocks.  At a resolution of $512^3$, this steepening scale is a few times smaller than the sonic length $l_s \sim 24 \Delta x$ where $v(l_s)=c_s$ or $P_v(l_s)=c_s^2=.035\mathcal{V}^2$.  Below the sonic length scale, turbulent velocity fluctuations become subsonic (and therefore less compressible).

\subsection{Discussion of Dynamical Results}
\begin{figure}
\plotone{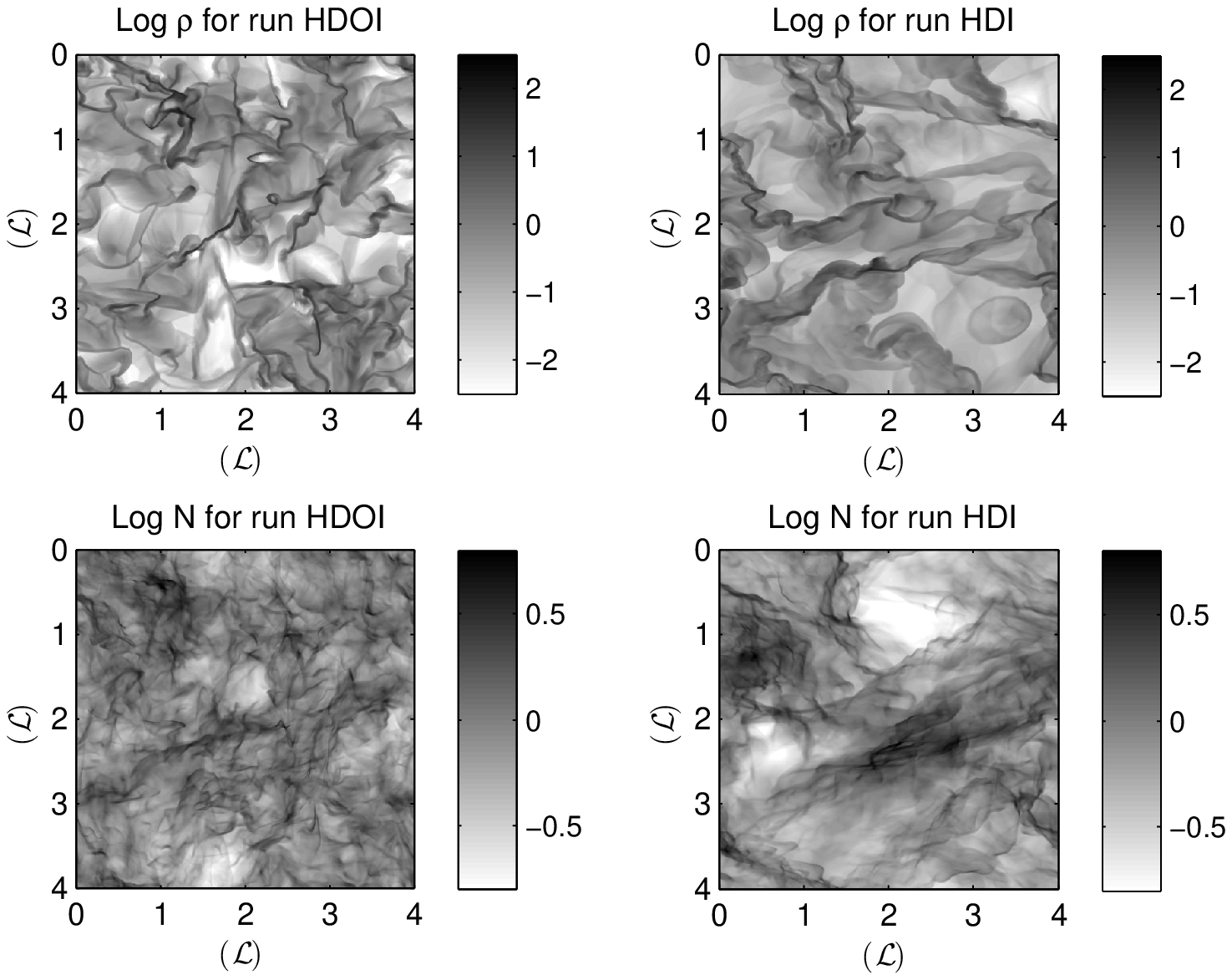}
\caption{Plots showing a slice through log density (top) and log column density (bottom) for run HDOI in which turbulence is driven by outflows and isotropic forcing (left panels) and for run HDI in which turbulence is driven by isotropic forcing only (right panels).  The density plots are scaled to $\rho_{_0}$, and the column density plots to $N_0=\rho_{_0}L_{box}$.  The x and y axis are in units of the outflow length scale $\mathcal{L}$}
\label{log_density}
\end{figure}

  Outflow driven turbulence is very different from isotropically forced turbulence, and the presence of outflows within an istropically driven cascade significantly alters the dynamics.  In general the presence of outflows will lead to a steeper velocity spectrum at sub-outflow scales and a shallower density spectrum at supra-outflow scales.  The steepening of the velocity spectrum at sub-outflow scales can be attributed to the presence of coherent outflows themselves, as well as their ability to effectively impede the cascade of energy by sweeping up small eddies as they expand.  As a result, the dynamics on sub-outflow scales are dominated by the passing of subsequent outflows rather than the cascading of energy from larger scales.  
\paragraph{}
The flatter density spectra at supra-outflow scales can be attributed to the ability of the outflows to disrupt larger scale structures from forming.  This can be most clearly seen upon visual inspection of the bottom panels in figure \ref{log_density}.  When the outflows are highly collimated (as was the case here), they can also disrupt the coherent density structures produced by neighboring outflows leading to a flattened density spectra even into the sub-outflow scale as seen in figure \ref{force_comp}.  Outflow feedback is therefore important in not only sustaining turbulence, but also in limiting the spatial scales of density structures and therefore shaping the IMF (initial mass function).  This interplay will be somewhat modified when the outflow locations and orientations are linked to the global dynamics of the clouds via gravitational collapse and magnetic fields, but we see as in \cite{Wang2010} that outflows have the potential to play a large role in setting the star formation rate and the initial mass function in their neighborhoods.
\paragraph{}
We note that our results do not imply that outflows can drive turbulence across the entire cloud.  Given the paucity of YSO outflows across the entire length of a cloud like Perseus it seems unlikely that the cloud as a whole can be set into turbulent motions from outflow feedback.  But this does not preclude outflow driving in regions where stars actually form. {\it The pervasive clustering inherent to star formation implies that outflows can significantly affect the star formation rate even though they likely do not drive the global GMC turbulence.}
\section{Connecting to Observations}
 Although the model of outflow driven turbulence used here is somewhat idealized by the outflows having uniform strength, collimation, and mean separation, it is reasonable to expect a similar signal to be present in the actual velocity spectra for regions where outflows are dynamically important.  We note that such a signal is likely to be blurred since outflows will have varying strength, collimation, and separation.  Even under the best of conditions, observers are still left with the difficult task of inferring the velocity spectra using VCA (velocity channal analysis), VCS, or PCA from data cubes of emission intensity as a function of position on the sky and line of sight velocity $T(x,y,v_z)$.  Since emission is at least density dependent, some assumptions on density-velocity correlations and distributions must be made.  These assumptions are often justified on the basis of both theoretical and numerical models of isotropically forced turbulence.  Figure \ref{2Dpdf} however, shows clearly that the density-velocity distributions and to some extent correlations are different between outflow driven turbulence and isotropically forced turbulence.  It is therefore worth investigating the ability of various techniques to reconstruct the velocity spectrum of models of outflow driven turbulence (as well as more realistic simulations that involve self-gravity etc.).  Here we look at the efficacy of one such technique (PCA) to infer the presence of outflows.
\subsection{Synthetic Observations}
  In practice, the actual emission from molecular clouds will depend on more than just density (ie temperature and chemistry).  In addition, optical depth effects further complicate the issue.  Here we ignore such effects and assume ideal observations of a density weighted velocity field.  This allows us to focus more directly on how the different density-velocity correlations and distributions produced by isotropic and outflow forcing effect the results inferred from PCA.  To convert our computational box into a spherical cloud we first tiled the box in each direction.  This allowed us to apply a spherical window as in \cite{bruntheyermaclow} without discarding as much information.  This also increased the ratio of cloud size to isotropic forcing scale and prevented a sort of blending of the first two principal components (described below).  We then created synthetic data cubes created by making density weighted histograms of the line of sight velocity along the x-axis at each yz position for runs HDI, HDOI, and HDO.  
\subsection{Global Line Profiles}
\begin{figure}
\plotone{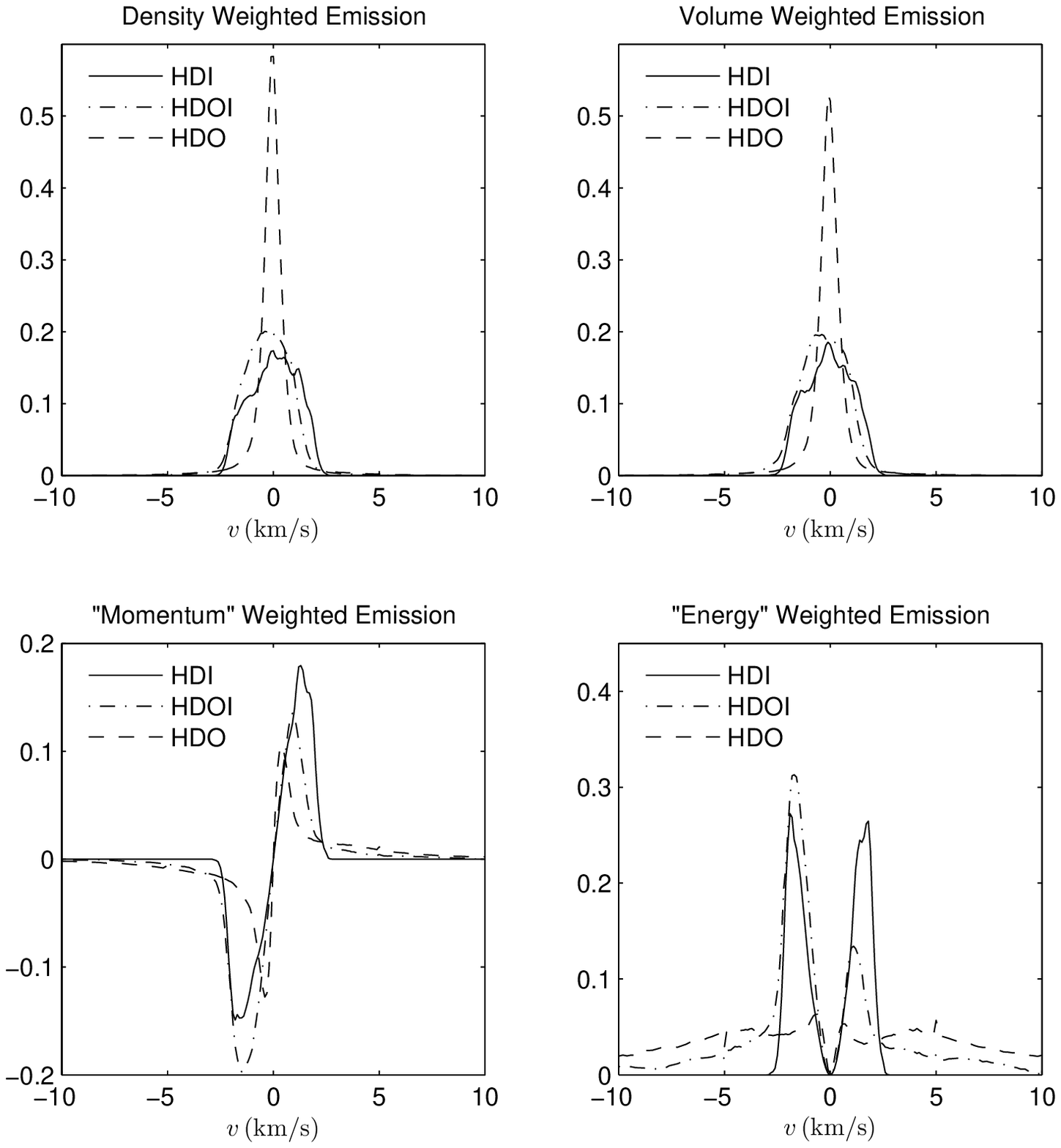}
\caption{Histograms of line of sight velocity for runs HDI, HDOI, \& HDO weighted by density $\rho$ (upper left), by volume/unweighted (upper right), by line of sight momentum $\rho v_z$ (lower left), and by $\rho v_z^2$ (lower right).  Note the high velocity components present in runs with outflow forcing (HDOI and HDO) that contribute to the momentum and energy, but produce a very small signal in the density weighted emission.}
\label{int_spect}
\end{figure}

  Before applying PCA it is instructive to examine the 'unresolved' spectra (integrated over the plane of the sky) for each run (upper left panel of figure \ref{int_spect}).  Note that runs with isotropic forcing (HDI and HDOI) have a broader central line profile then the run with just outflow forcing (HDO) roughly consistent with the greater velocity scale $\mathcal{V}_I=2\mathcal{V}$ associated with the isotropic forcing as well as the saturation levels of momentum seen in the left panel of figure \ref{totals}.  It is not surprising that the linewidths are more consistent with the saturation levels of momentum as opposed to velocity dispersions since both momentum and emission are weighted by density.  
\paragraph{}
Additionally the line profiles for runs with outflow forcing (HDO and HDOI) have extended wings not present in run HDI due to the higher velocity material injected by the outflows.  While the wings are only marginally apparent in the density weighted emission and as such do not contribute to the observed line-widths, they do contribute significantly to the momentum and energy.  This becomes apparent after weighting the 'observed' line profiles by the line of sight velocity $v_z$ or by $v_z^2$.  These modified spectra seen in the bottom panels of figure \ref{int_spect} show the distribution of momentum or energy at the different velocities.  Note the large contribution to the momentum and energy provided by the high velocity wings in runs HDO and HDOI.  As a result, line-widths inferred from density weighted emission will tend to underestimate the energy and momentum injected by the outflows.  In addition to the line profiles produced by density weighted emission, we constructed line profiles of volume (or unweighted) emission (upper right panel of figure \ref{int_spect}).  Note the global line profiles are very similar implying a lack of any strong density velocity correlations.  This is not surprising since the density-velocity distributions in figure \ref{2Dpdf} are only weakly correlated over the bulk of the material for all three runs.

\subsection{Principal Component Analysis}
\begin{figure}
\plotone{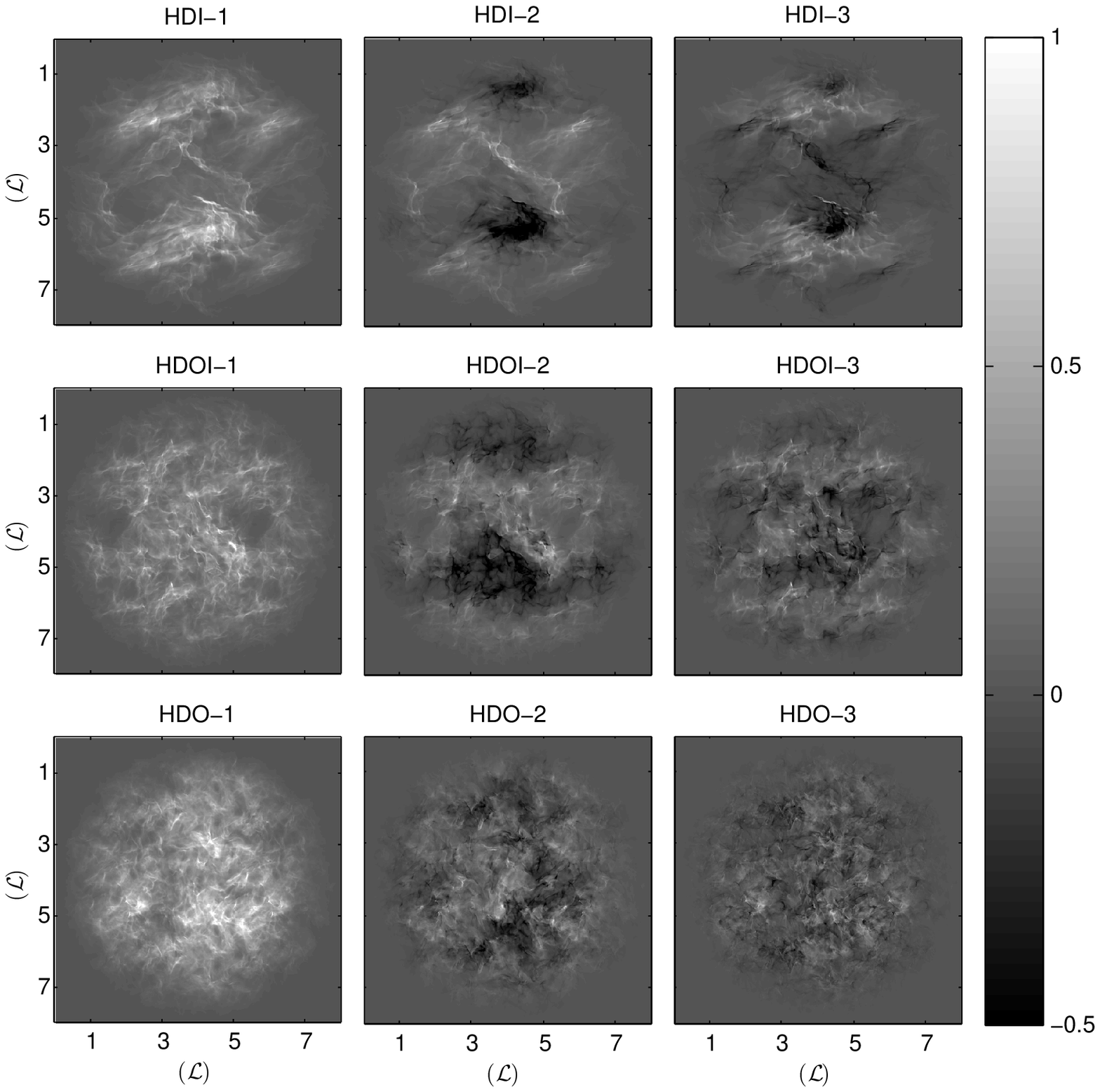}
\caption{First three principal eigenimages constructed from PCA analysis of synthetic data cubes (left to right) of runs HDI (top), HDOI (middle), and HDO (bottom).  All eigenimages are normalized to mean emission (column density).  Note the lack of large scale coherent structures in the first principal eigenimage for runs HDOI and HDO where outflows are present, and the large dipole structure present for runs HDOI and HDI where large scale forcing is present.}
\label{pcomp1}
\end{figure}

\begin{figure}
\plotone{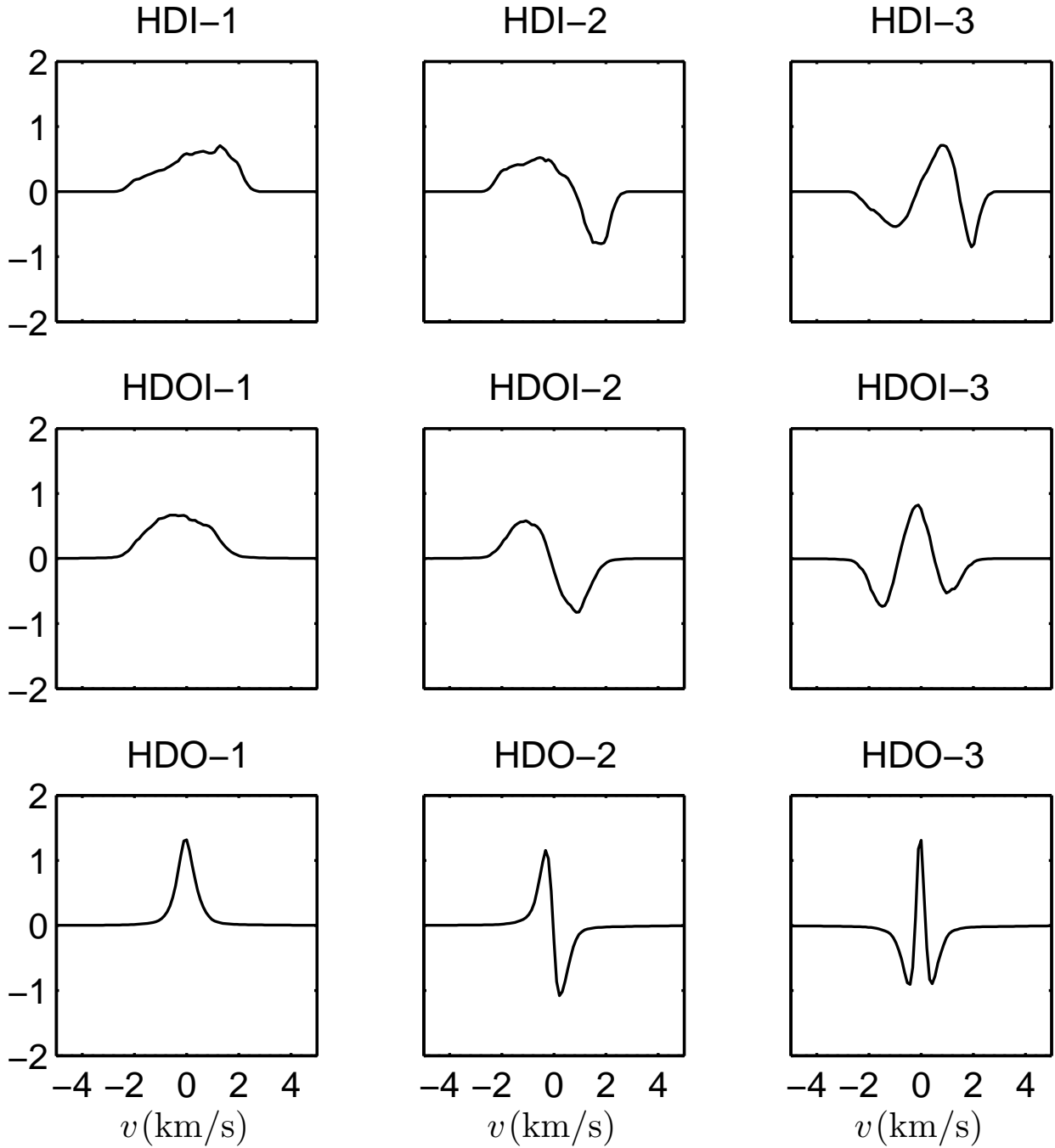}
\caption{First three principal eigenvectors constructed from PCA analysis of synthetic data cubes (left to right) of runs HDI (top), HDOI (middle), and HDO (bottom).  Note the broader eigenvectors for runs HDOI and HDI where isotropic forcing is present compared to run HDO where only outflow forcing is present.  All eigenvectors are normalized to unity.}
\label{pcomp2}
\end{figure}

\begin{figure}
\plotone{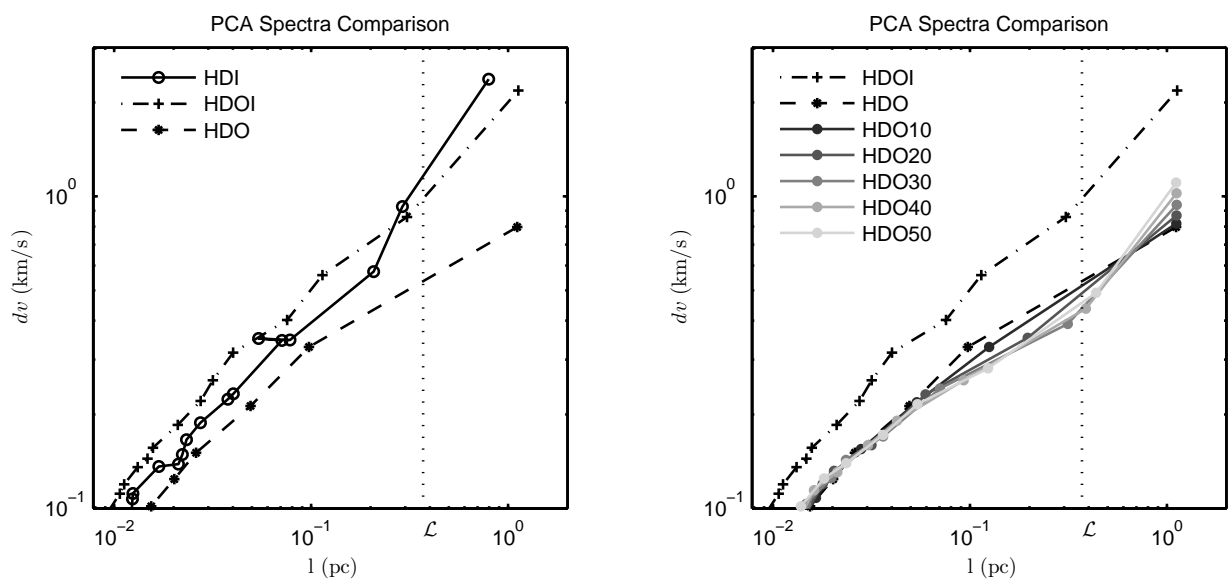}
\caption{Plots showing the line-widths and correlation lengths for the spectroscopically resolved principal components.  Left panel shows the results for runs HDI, HDOI, and HDO.  Right panel shows how the line-widths and correlation lengths change when a sinusoidal perturbation of varying strength is added to the final frame of run HDO.}
\label{pcaspect2}
\end{figure}

\begin{figure}
\plotone{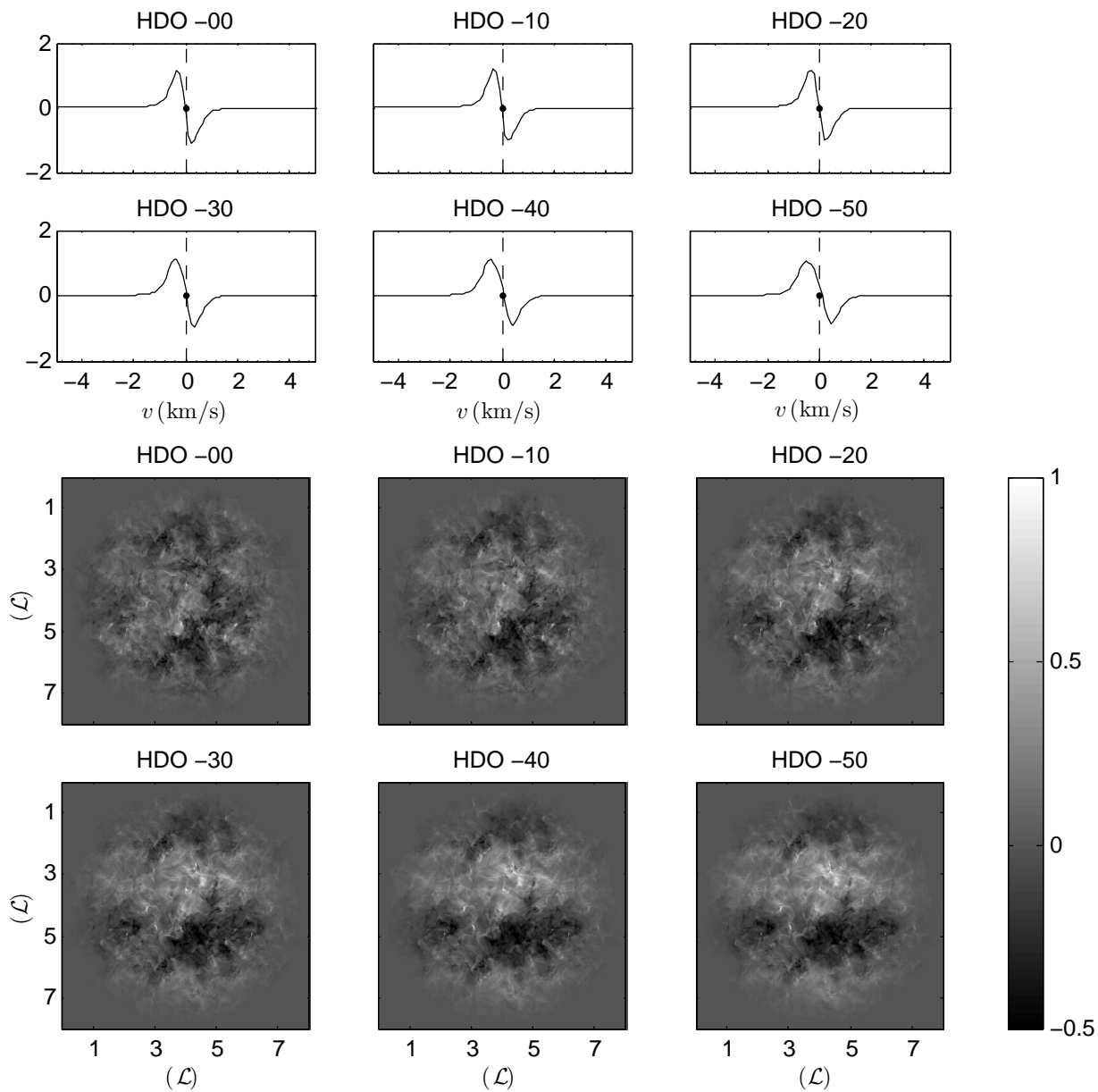}
\caption{Plot showing how the second principal eigenspectra (top) and eigenimage (bottom) change when a sinusoidal perturbation of varying strength is added to the final frame of run HDO.  Plots are scaled as in figures \ref{pcomp1} and \ref{pcomp2}.  Note the slight broadening of the eigenspectra and the appearance of a large dipole structure in the eigenimages resulting in a larger correlation length ($l_2$)}
\label{pcomp3}
\end{figure}

  Principal Component Analysis is a technique for finding relationships between a given set of variables that can best explain the variance in a given set of measurements.  Given a set of $n$ measurements of $p$ variables $T_{ij}$ (corresponding to the $i^{th}$ measurement of variable $j$), PCA first involves the construction of the covariance matrix $S_{jk}=\displaystyle\sum\limits_{i=1}^n{\hat{T}_{ij}\hat{T}_{ik}}$ where we have subtracted off the mean values for each variable ($\hat{T}_{ij}=T_{ij}-\bar{T}_j$ where $\bar{T}_j=\frac{1}{n}\displaystyle\sum_{i=1}^n{T_{ij}}$).  The first principal component is then the (normalized) eigenvector $V^1_j$ of the covariance matrix (and corresponding projection $P^1_i=\displaystyle\sum\limits_{j=1}^p{\hat{T}_{ij}V^1_j}$) with the largest eigenvalue $\lambda_i$ and as such can account for the largest source of variance.  The second principal component is the eigenvector with the second largest eigenvalue and so on.
\paragraph{}
  Principal Component Analysis as applied to interstellar turbulence is formally described in \cite{pca} and \cite{bruntheyera}.  It involves treating the spectroscopic datacube $T_{ij}=T(x_i,y_i,v_j)$ as a measurement at each pixel $(x_i,y_i)$ of a set of variables (the emission in each velocity channel $v_j$).  The eigenvectors of the covariance matrix are then line spectra in velocity space, and the projections are images in $xy$ space where each pixel represents the degree of correlation between the line spectra at that pixel and the principal eigenvector.  These principal 'eigenspectra' and corresponding 'eigenimages' then account for the largest sources of variance in the datacube.  While PCA formally uses the covariance matrix, a modified version (mPCA) has also been used to decompose datacubes into eigenspectra and eigenimages - though using the 'co-emission' matrix $\tilde{S_{jk}}=\displaystyle\sum_{i=1}^n{T_{ij}T_{ik}}$ in which the mean values for each channel are not subtracted.  The $ij^{th}$ element of the co-emission matrix is therefore the degree of correlation between the emission in velocity channels $i$ and $j$ summed over all positions. The lack of any mean value subtraction introduces additional pseudo-variance which is effectively removed in the first principal component.  Here we use mPCA to make contact with previous work \citep{bruntheyera, bruntheyermaclow, bruntheyera}
\paragraph{}
  PCA effectively reduces the dimensionality of the dataset by projecting the 3-D datacube onto a sequence of 1-D principal eigenspectra resulting in a sequence of 2-D principal eigenimages.  The generation of the eigenspectra (and therefore the eigenimages) however, does not make use of the velocities associated with each channel (or the positions on the sky of each pixel).  The positions in the sky of each spectra could be shuffled without changing the eigenvectors, and the velocity channels could be shuffled without changing the resulting eigenimages.  However, the resulting (unshuffled) eigenvectors and eigenimages can be used to determine sequentially the velocity magnitudes ($dv_i$) and length scales ($l_i$) over which the largest sources of variance change.  These scales are typically calculated as the distance at which the autocorrelation functions of the eigenspectra ($dv_i$) and eigenimages ($l_i$) drop by a factor of $e$.  The combination of $l_i$ and $dv_i$ for each principal component then allow for the reconstruction of a line-width size relationship (or equivalently a velocity spectra $\mathcal{E}_{PCA}$).  This inferred spectra has been shown to closely mimic the actual velocity spectra in fractional Brownian motion simulations \citep{bruntheyera}.  In addition, \citep{bruntheyermaclow} demonstrated that the ratio of $l_2$ to $l_1$ (when using mPCA) is correlated with the ratio of cloud size $\lambda_D$ to driving scale $L_{box}$  in simulations of isotropically forced turbulence.  Since the first principal component in mPCA is likely to mirror the mean values to account for the pseudo-variance, the principal length $l_1$ is typically the size of the cloud.  This leaves the second principal component and corresponding length $l_2$ to account for the largest source of true variance which should be associated with the driving scale.  
\paragraph{}
Figures \ref{pcomp1} and \ref{pcomp2} show the first three eigenimages and eigenvectors for the three different cases.  Note that while all three runs have similar velocity dispersions (figure \ref{totals}), the runs with isotropic forcing (HDI and HDOI) have much broader eigenspectra (figure \ref{pcomp2}).  This broadening results in higher spectral correlation magnitudes $dv_i$ for runs HDI and HDOI given in table \ref{autocorrs}.  These differences however, are in general consistent with the differences seen in the final values of the mean scalar momentum in figure \ref{totals}, and is not surprising considering that both the mean scalar momentum and emission are density weighted.  It is also interesting to note that the first three spectral correlation lengths for run HDOI are actually smaller ($\approx 10\%$) than those for run HDI even though run HDOI has additional forcing via outflows, a larger velocity dispersion, and a larger saturation value of mean scalar momentum.  We interpret this decrease in velocity correlation length as arising from the disruption by outflows of the large scale coherent flows produced by the isotropic forcing.  This is consistent with the decrease in power at large scales seen in the velocity spectra in figure \ref{force_comp}.  
\paragraph{}
While the presence of outflows in run HDOI does not contribute to larger velocity correlation lengths ($dv_1$, $dv_2$, \& $dv_3$) and as such is not readily apparent in the principal eigenspectrum, the presence of outflows is apparent in the first principal eigenimage in the lack of coherent large scale density structures for run HDOI compared to run HDI.  This is consistent with the flattening of the density spectra seen in figure \ref{force_comp} and the larger principal autocorrelation length ($l_1$) given in table \ref{autocorrs}.  However, the second and third eigenimages for run HDOI are very similar to run HDI and show clear evidence for the large scale driving.  These large dipole structures present in the second eigenimages for runs HDI and HDOI result in larger autocorrelation lengths ($l_2$) and larger ratios of $l_2$ to $l_1$ especially when compared to run HDO (see table \ref{autocorrs}).  We find as in \citep{bruntheyermaclow} that the ratio of $l_2$ to $l_1$ correctly identifies the largest driving scale.  Runs HDI and HDOI are driven isotropically with $\frac{\lambda_D}{L_{box}}=1$ and have ratios $\frac{l_2}{l_1}=.3649$ and $.2723$ respectively, while run HDO is driven by outflows at one fourth the scale ($\frac{\lambda_D}{L_{box}}=.25$) and has a corresponding ratio roughly one fourth ($\frac{l_2}{l_1}=.0874$.  While the ratio of $l_2$ to $l_1$ does not reveal the presence of the outflows in run HDOI, the resulting line-width size relationships shown in the left panel of figure \ref{pcaspect2} do show an excess of energy for run HDOI beginning at the third principal component whose correlation length $l_3=.1142$ is approximately the same as that for the second principal component for run HDO $l_2=.0973$ which is the scale associated with the outflow driving.  We note however, that since the inferred spectra from PCA relies on line widths that are fairly insensitive to the high velocity wings, PCA will in general underestimate the true strength of the outflow forcing.  This explains why run HDI appears to have more power over the entire range of scales in the results from PCA when in fact run HDI has less power at small scales as seen in the true velocity spectrum in figure \ref{force_comp}.
\paragraph{}
We have already seen how the density weighting of the emission results in linewidths consistent with values of the mean scalar momentum rather than the velocity dispersion.  This alone explains why the second principal component for run HDOI picks out the variance produced by the isotropic forcing rather then that due to the ouflows.  However, since the emission is not weighted by velocity, low velocity motions on a given scale can produce just as much variance in the datacube as high velocity signals on the same scale. {\it Thus a low velocity large scale coherent motion could produce more variance than a high velocity motion of less coherence on that scale.  Thus the ratio of $l_2$ to $l_1$ is more a simple measure of the largest scale of coherent motion rather than the scale of dominant driving energy.}  In simulations of isotropically forced turbulence there is no distinction.  But in clouds where outflows are present within an external cascade, or where there are large scale coherent motions due to angular momentum conserving collapse, this distinction becomes important.  
\paragraph{}
To further test the aforementioned bias we ran PCA on a series of datacubes produced by modifying the velocity field in run HDO through the addition of a large scale sinusoidal variation in the line of sight velocity - similar to what is seen in run HDOI.  We varied the strength of the perturbation from $10\% \mathcal{V}$ to $50\% \mathcal{V}$ in $10\%$ increments (HDO10, HDO20, HDO30, HDO40, \& HDO50) and looked at how the resulting correlation lengths shifted.  The first three eigenimages for each run are shown in figure \ref{pcomp3}.  By run HDO30, the second principal eigenimage shows the large scale velocity variation and the resulting autocorrelation length $l_2$ has increased from its initial value of $.0587$ to $.1477$ corresponding to the driving scale seen in runs HDI and HDOI.  Note also the (small) $20\%$ increase in $dv_1$ from $.10245$ to $1.2009$ produced by the velocity perturbation.  These trends are also evident in the line-width size relationships shown in the right panel of figure \ref{pcaspect2}.  Thus our results indicate that PCA is not a reliable measure for determining the specific question of the dominant mode of energy input in driving turbulence.  In our experiments even though the energy put into the solenoidal motions was smaller than that in the outflows, PCA still picked up these solenoidal motions as dominant i.e. $E_{sol} < E_{of}$ but $l_{2(sol)} > l_{2(of)}$.
\begin{deluxetable}{l c c c c c c c c c}
\tablecolumns{10}
\tablewidth{0pt}
\tablecaption{Autocorrelation lengths of eigenimages and eigenvectors from principal component analysis.  All lengths are in pc and velocities in km/s\label{autocorrs}}
\tablehead{\colhead{Run} & \colhead{$l_1$} & \colhead{$l_2$} & \colhead{$l_3$} & \colhead{$l_2/l_1$} & \colhead{$\Lambda_D/L_{box}$} & \colhead{$dv_1$} & \colhead{$dv_2$} & \colhead{$v_{rms}$} & \colhead{$\frac{dv_1}{v_{rms}}$} }
\startdata
HDI   & .7960 & .2904 & .2076 & .3649 & 1.000 & 2.376 & .9276 & 1.9785 & 1.201 \\
HDOI  & 1.126 & .3067 & .1142 & .2723 & 1.000 & 2.183 & .8593 & 2.2546 & .9682 \\
HDO   & 1.113 & .0973 & .0492 & .0874 & .2500 & .7977 & .3293 & 2.2907 & .3482 \\
HDO10 & 1.115 & .1251 & .0537 & .1123 & \nodata{} & .8187 & .3289 & \nodata{} & \nodata{} \\
HDO20 & 1.117 & .1964 & .0595 & .1758 & \nodata{} & .8684 & .3520 & \nodata{} & \nodata{} \\
HDO30 & 1.119 & .3138 & .0703 & .2804 & \nodata{} & .9402 & .3903 & \nodata{} & \nodata{} \\
HDO40 & 1.118 & .3874 & .0927 & .3466 & \nodata{} & 1.023 & .4370 & \nodata{} & \nodata{} \\
HDO50 & 1.112 & .4367 & .1235 & .3928 & \nodata{} & 1.108 & .4907 & \nodata{} & \nodata{} \\
\enddata
\end{deluxetable}
\section{Conclusion} \label{conclusion}
Building on the work of \cite{Carroll2009} we studied the properties of turbulence driven isotropically, by outflows, and by a combination of both for input parameters typical of clump scale molecular clouds.  We find as in \cite{Carroll2009} that outflows are able to drive supersonic turbulence at levels consistent with the scaling relations put forth by \cite{matzner2007}. 
We also find that turbulence driven by outflows is characterized by a steeper velocity spectrum at sub-outflow scales, and an overall flatter density spectrum.  Based on our resolution study, we conclude that when outflows are present, the resulting steep velocity spectra and correspondingly smooth velocity structures seen at sub-outflow scales are not due to numerical dissipation, but rather due to the presence of the outflows themselves.  Outflows are able to continually generate smooth velocity structures on outflow scales that sweep up and 'dissipate' smaller structures at the same rate that these structures would begin to cascade.  Outflow driven turbulence is therefore quite different from the cascade of energy produced by large scale isotropic forcing.
\paragraph{}
Since outflows drive motions on sub-parsec scales, it seems likely that realistic outflow driving will take place in the presence of a externally driven cascade, due to external energy inputs into molecular cloud systems from much larger scales.  With this motivation, we drove outflows into a medium also supplied with isotropcially driven turbulence from larger scales.  We found that the properties of the turbulence at sub-outflow scales were largely unchanged by the external isotropic cascade for typical cloud parameters. In particular, the velocity spectra still contain a knee at the outflow scale and a steep slope at sub-outflow scales.  We also found that the outflows produced a flatter density spectra even at supra-outflow scales.  We attribute this to the disruption by outflows of larger scale structures produced by the large scale driving.  We conclude therefore, as in \cite{Wang2010}, that outflows have the potential to play a large role in setting the star formation rate and the initial mass function in their neighborhoods.
\paragraph{}
 Finally, we also find that the density-velocity distributions and to a smaller extent correlations present in outflow driven turbulence are quite different from those present in an isotropic cascade.  Many of the techniques for inferring the velocity spectrum of turbulence in star forming regions rely on assumptions about density-velocity distributions and correlations based on simulations of isotropically forced turbulence.  To illustrate the resulting potential bias, we  tested one such model (PCA) against the results from our simulations of outflow driven turbulence.  We find that indeed PCA underestimates the contribution of outflows to the velocity dispersion, though not necessarily the mean scalar momentum.  We find that PCA is able to infer the driving scale of the outflows for the case of pure outflow driven turbulence, but that the presence of even small velocity large scale flows quickly leads to an overestimation.  We suggest weighting the data cubes by the line of sight velocity as one way to make PCA more sensitive to higher velocity flows.  This highlights that nature of the forcing of turbulence (e.g. isotropic on large scales vs. outflow driven) is of direct importance for interpreting observations.
\acknowledgments{}

Support for this work was in part provided by NASA through awards issued by
JPL/Caltech through Spitzer program 20269, the Department of Energy through
grant number DE-SC-0001063, Cornell University through agreement number
41843-7012, the National Science Foundation through grants AST-0807363 as
well as the Space Telescope Science Institute through grants HST-AR-11250
and HST-AR-11251.  We also thank the University of Rochester Laboratory for
Laser Energetics and funds received through the DOE Cooperative Agreement
No. DE-FC03-02NA00057.

This research was also supported in part by the Center for Research Computing at the University of Rochester as well as the National Science Foundation through TeraGrid resources provided by the National Center for Supercomputing Applications, the San Diego Supercomputer Center, \& the Pittsburgh Supercomputing Center under grant number TG-AST080035T.

\end{document}